\documentclass{blois}

\usepackage{amsmath,amssymb}

\newcommand{\Photo}{}

\begin{document}
\vspace*{4cm}
\title{Theory status of hadronic top-quark pair production}

\author{Christian Schwinn~\footnote{Supported by the Heisenberg Programme of the DFG}}

\address{Institut f\"ur Theoretische Teilchenphysik und   
Kosmologie,  \\
RWTH Aachen University, Sommerfeldstra\ss e 16,  D--52056 Aachen, Germany}

\maketitle\abstracts{The status of theoretical predictions for top-quark pair
  production at hadron colliders is reviewed, focusing on the total cross
  section, differential distributions, and the description of top-quark
  production and decay including off-shell effects.}

\section{Introduction}
The increasingly precise measurements of top-quark observables at the LHC
require corresponding progress in their theoretical description. This
contribution provides an overview of recent theory developments, from
corrections beyond next-to-next-to-leading order~(NNLO) for the total cross
section in Section~\ref{sec:total}, via state-of-the-art combinations of NNLO
QCD and NLO electroweak corrections for differential observables for stable
tops in Section~\ref{sec:differential}, to descriptions of the full production
and decay process and the effects on the $m_t$ measurement in
Section~\ref{sec:decay}.

\section{Total cross section}
\label{sec:total}
\begin{figure}[t]
  \centering
  \includegraphics[width=.48\linewidth]{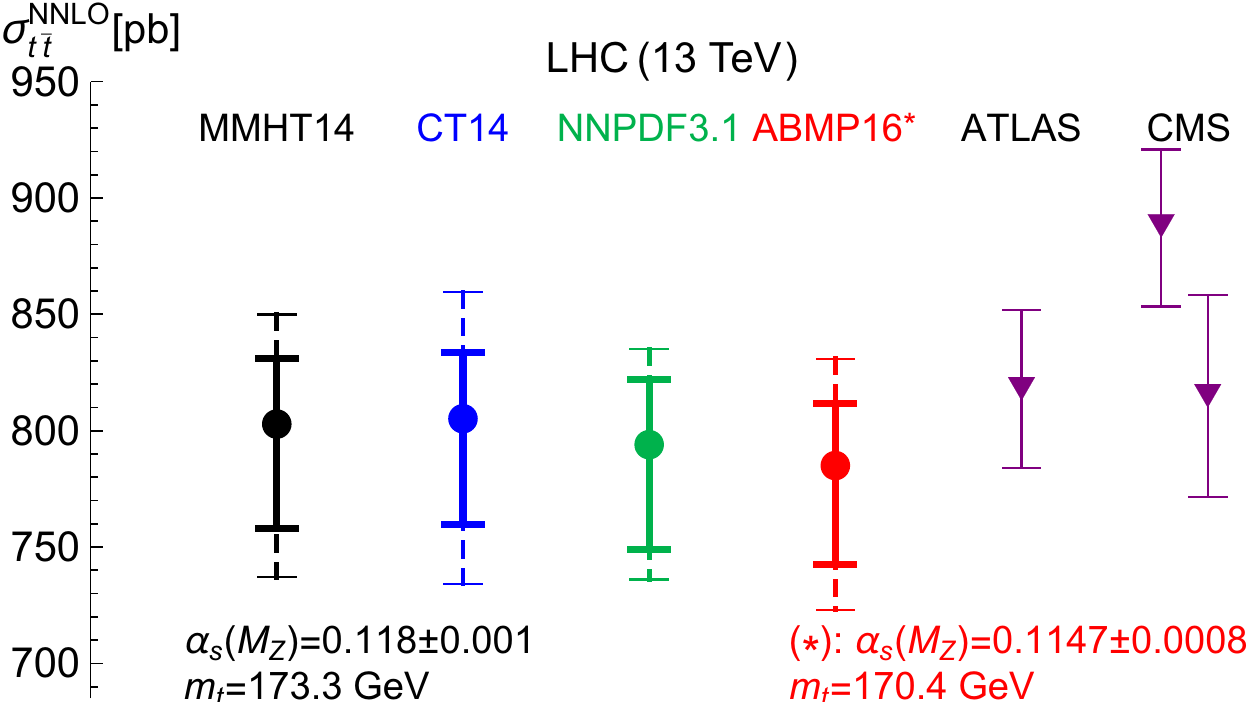} \hfill
  \includegraphics[width=.48\linewidth]{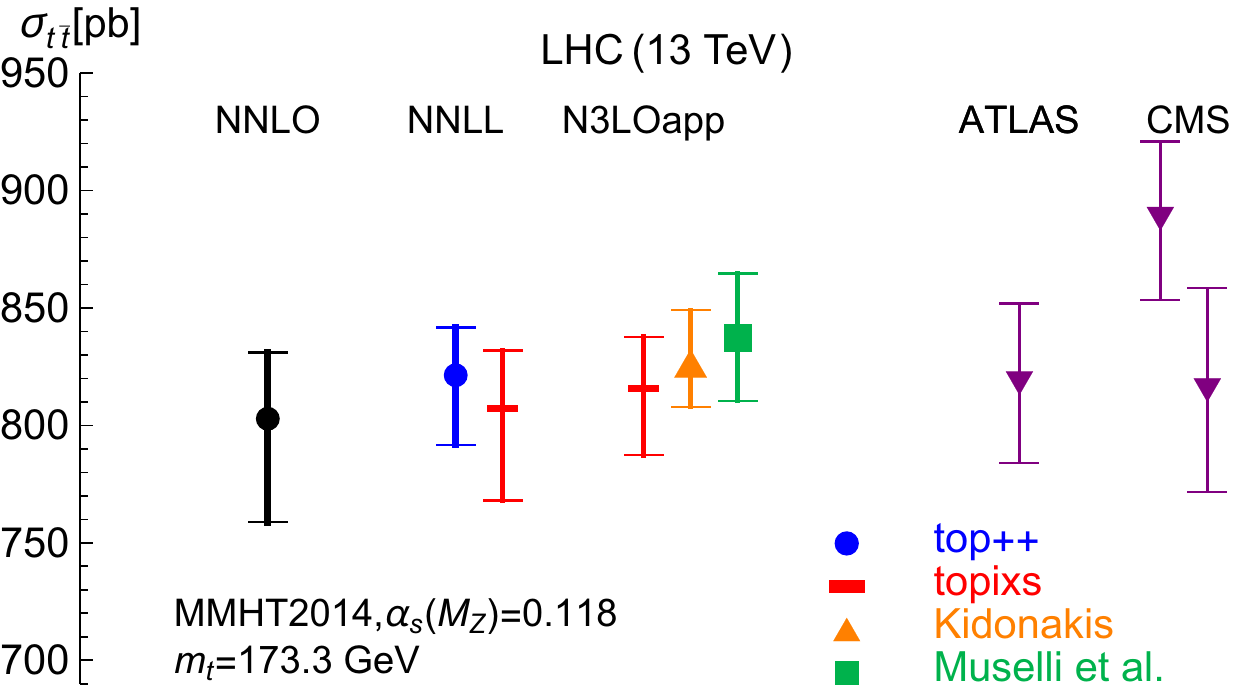}
  \caption{Left: Comparison of the total $t\bar t$ cross section at NNLO QCD
    with different PDFs to LHC measurements at $13$~TeV including the scale
    uncertainty (solid lines) and the $68\%$ confidence-level PDF+$\alpha_s$
    uncertainty (dashed lines). Right: Different predictions of higher-order
    threshold corrections. For \texttt{topixs} and Muselli et al.\ the scale
    uncertainty is added in quadrature to the systematic uncertainty of the
    threshold approximation. The relative corrections of Muselli et al.\ and
    Kidonakis have been rescaled by the NNLO result with MMHT2014
    PDFs.}\label{fig:sigtotal}
\end{figure}

The total top-quark pair production cross section provides an important
benchmark measurement at the Tevatron and LHC that allows to measure the top
mass in a well-defined scheme with an accuracy of
$2$~GeV,\cite{Aad:2014kva,Khachatryan:2016mqs} and to determine the strong
coupling constant $\alpha_s$ with an accuracy competitive with other
determinations.\cite{Klijnsma:2017eqp,Chatrchyan:2013haa} A comparison of the
NNLO QCD prediction~\cite{Czakon:2013goa} for different PDF
sets~\cite{Harland-Lang:2014zoa,Dulat:2015mca,Ball:2017nwa,Alekhin:2017kpj} to
the most precise experimental LHC results at
$\sqrt{s}=13$~TeV~\cite{Aaboud:2016pbd} in Figure~\ref{fig:sigtotal} shows
good agreement among the PDF sets~\footnote{ For ABMP16~\cite{Alekhin:2017kpj}
  the values for $\alpha_s(M_Z)$ and $m_t$ preferred by the PDF fit have been
  employed. Note that in this set a well as in
  MMHT14~\cite{Harland-Lang:2014zoa} and NNPDF3.1,\cite{Ball:2017nwa} LHC data
  on $t\bar t$ production are already used in the fit.}  and with the
experimental results, whose precision challenge that of the NNLO calculation.
Since a full N$^3$LO calculation is currently out of reach, attempts to reduce
the perturbative uncertainties further rely on resummation methods to compute
higher-order corrections enhanced in certain kinematic limits.  For the total
partonic cross section, corrections enhanced in the partonic threshold limit
$\beta=\sqrt{1-4 m_t^2/\hat s}\to 0$ are given by logarithmic soft-gluon
corrections, $\alpha_s \ln^2\beta$, and Coulomb corrections, $\alpha_s/\beta$.
The logarithmic accuracy of the resummed cross section can be defined in
a``primed'' and ``unprimed'' counting, depending on the inclusion of
fixed-order non-logarithmic corrections:
\begin{align}
\label{eq:syst}
\hat{\sigma} &\propto \hat\sigma^{(0)} 
\sum_{k=0} \left(\frac{\alpha_s}{\beta}\right)^k \!\!\!
\exp\Big[\underbrace{\ln\beta\,g_0(\alpha_s\ln\beta)}_{\mbox{(LL)}}+ 
\underbrace{g_1(\alpha_s\ln\beta)}_{\mbox{(NLL)}}+
\underbrace{\alpha_s g_2(\alpha_s\ln\beta)}_{\mbox{(NNLL)}}+
\underbrace{\alpha_s^2 g_3(\alpha_s\ln\beta)}_{\mbox{(N$^3$LL)}}+\ldots\Big]
\nonumber\\[0.2cm]
& \quad\times
\left\{1\,\mbox{(LL,NLL)}; \alpha_s,\beta \,\mbox{(NLL',NNLL)}; 
\alpha_s^2,\alpha_s\beta,\beta^2 \,\mbox{(NNLL',N$^3$LL)};
\ldots\right\}.
\end{align}
In Figure~\ref{fig:sigtotal} the \texttt{top++ 2.0}~\cite{Czakon:2011xx}
result for NNLL soft-gluon resummation in Mellin space~\cite{Cacciari:2011hy}
is compared to the combined soft/Coulomb-gluon resummation in momentum
space~\cite{Beneke:2011mq} implemented in \texttt{topixs}.\cite{Beneke:2012wb}
The main numerical difference is due~\cite{Piclum:2018ndt} to an NNLL' effect
from $\mathcal{O}(\alpha_s^2)$ constants included in \texttt{top++}, while the
Coulomb corrections included in \texttt{topixs} have a smaller effect.

As an alternative to all-order resummation, approximate N$^3$LO results have
been computed using different resummation methods and logarithmic
accuracy. The calculation of Kidonakis~\cite{Kidonakis:2014isa} is based on
NNLL resummation for one-particle inclusive kinematics.  Muselli et
al.\cite{Muselli:2015kba} use partial N$^3$LL soft resummation in Mellin
space, including $1/N$-suppressed contributions and information on the
large-$\beta$ behaviour.  A calculation based on partial N$^3$LL soft
resummation and N$^3$LO Coulomb corrections~\cite{Piclum:2018ndt} is
implemented in \texttt{topixs3.0}.
All N$^3$LO${}_{\text{app}}$ predictions are consistent with the resummed NNLL
calculations and have a similar scale uncertainty of $\pm 3\%$, which suggests
that corrections beyond N$^3$LO are indeed small.  The range of the different
implementations of higher-order threshold corrections, which include
complementary effects, indicates that the scale uncertainty underestimates the
true perturbative uncertainty and should be supplemented by an $1$--$2\%$
estimate of systematic
uncertainties\cite{Beneke:2011mq,Muselli:2015kba,Piclum:2018ndt} due to the
threshold approximation.

\section{Differential distributions}
\label{sec:differential}
With the increase of the LHC centre-of-mass energy to $13$~TeV and
accumulating luminosity, the measurement of observables such as
transverse-momentum or invariant-mass distributions in the TeV range becomes
increasingly precise and requires matching theoretical predictions.  The
theoretical description of such differential distribution has recently been
achieved at NNLO accuracy in QCD.\cite{Czakon:2016ckf} As an application of
differential NNLO predictions, it has been advocated to constrain the gluon
PDF using rapidity distributions,\cite{Czakon:2016olj} which are argued to be
robust against variations of the top mass and possible new-physics effects.
For specific observables, fixed-order calculations can be complemented by NNLL
resummation in various kinematic regimes such as the low $p_T$
limit.\cite{Li:2013mia,Catani:2018mei} NNLL resummation for boosted
tops,\cite{Pecjak:2016nee} which simultaneously resums threshold logarithms in
the pair-invariant mass, $\ln(1-M^2_{t\bar t}/\hat s)$, and mass logarithms
$\ln(m_t^2/\hat s)$, has allowed to validate dynamical scale choices
$\mu_f=H_T/4 =\frac{1}{4}(\sqrt{m_t^2+p_{T,t \phantom{\bar
      t}}^2}+\sqrt{m_t^2+p_{T, \bar t}^2})$ for the $M_{t\bar t}$ spectrum and
$\mu_f=m_T/2$ for the $p_t$ spectrum.\cite{Czakon:2018nun}

At high $p_T$, both higher order QCD corrections and EW corrections, which are
dominated by Sudakov effects, are necessary for a proper theoretical
description. A multiplicative combination of NNLO QCD and NLO EW
corrections~\cite{Czakon:2017wor} is advocated to capture the dominant
higher-order soft-QCD/EW-Sudakow effects at large $p_T$.  In a complementary
work,\cite{Gutschow:2018tuk} the interplay of real-gluon emission and EW
corrections is modeled using parton-shower merging of the processes $t\bar t$
and $t\bar tj$ at NLO QCD$+$EW$_{\text{virt}}$ accuracy.  Here the NLO EW
corrections are included in a virtual approximation supplemented with real
photonic corrections in the YFS approximation.  The results of these works
show that the EW corrections become relevant for the comparison to
experimental data for $p_T>500$~GeV.

\section{Top-quark production and decay}
\label{sec:decay}
\begin{figure}[t]
  \begin{center} \includegraphics[width=.96\linewidth]{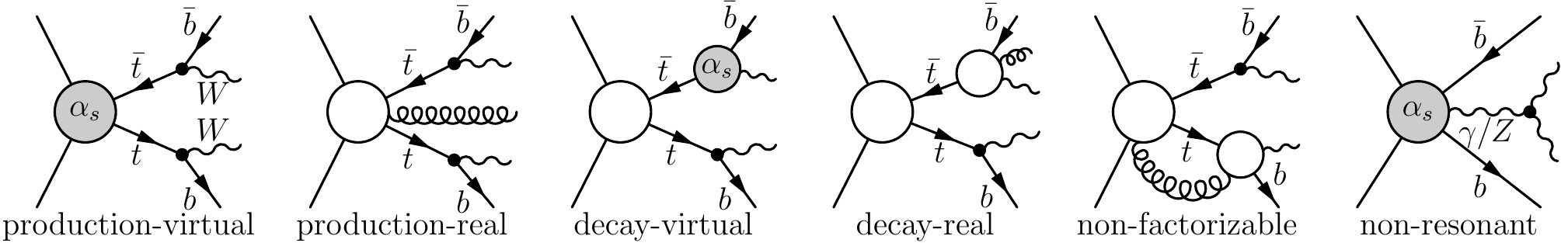}
  \end{center}
\caption{
  Example topologies of diagrams for the process $pp\to b\bar b W^+W^-$ arising in the narrow-width
  approximation, the double-pole approximation, and the full off-shell NLO QCD
  calculation for unstable top quarks. The decay products of the $W$ bosons are not shown.
}
\label{fig:off-shell}
\end{figure}

Kinematic observables entering studies of top-quark properties such as spin
correlations or the "direct" measurement of the top-quark mass require to go
beyond the approximation of stable top quarks and include the top decay
products.  Since the decay width of the top quark is relatively small,
$\Gamma_t/m_t\sim 1 \%$, in generic kinematic situations the narrow-width
approximation~(NWA) of the production cross section of the resulting
$ b\bar b+ 4f$-final state may be performed,
\begin{equation}
  \sigma_{pp\to b\bar b\, 4f}\Rightarrow \sigma_{pp\to t\bar t}\times
\frac{\Gamma_{t\to  b f_1f_2 }}{\Gamma_t}
  \frac{\Gamma_{\bar t\to \bar b f_3f_4}}{\Gamma_t}.
\end{equation}
In the framework of the NWA, spin correlations and
QCD~\cite{Bernreuther:2004jv,Melnikov:2009dn} as well as EW
corrections~\cite{Bernreuther:2010ny} can be incorporated, including real and
virtual corrections to the production and decay of on-shell tops.  In a step
beyond the NWA, the double-pole approximation additionally includes
Breit-Wigner propagators and non-factorizable corrections connecting initial
and final states.\cite{Falgari:2013gwa} NLO computations for the full
$pp\to b\bar b +4f$ processes further include full off-shell effects and
non-resonant contributions.  Here NLO QCD corrections for the di-lepton
$pp\to b\bar b
\ell\nu_\ell\ell'\nu_{\ell'}$~\cite{Denner:2010jp,Bevilacqua:2010qb} and
semileptonic final state $pp\to b\bar b \ell\nu_\ell jj$,\cite{Denner:2017kzu}
as well as EW corrections for the di-lepton final state~\cite{Denner:2016jyo}
are available.  Example NLO QCD topologies for the different contributions are
shown in Figure~\ref{fig:off-shell}.  A description of top production and
decay at NNLO is currently only feasible in the NWA, where an approximate
differential NNLO calculation of top-quark production~\cite{Broggio:2014yca}
has been combined~\cite{Gao:2017goi} with the NNLO decay
corrections.\cite{Gao:2012ja,Brucherseifer:2013iv} Recently also preliminary
results using the exact NNLO production cross section have been
presented.\cite{Poncelet:top2018} These calculations show promise for an
improved description of experimental results on spin correlations.

Comparisons of the NWA and fully off-shell NLO QCD
calculations~\cite{AlcarazMaestre:2012vp,Heinrich:2017bqp} show the expected
accuracy of $\mathcal{O}(\Gamma_t/m_t)$ for generic kinematics, provided NLO
decay corrections are included, which can reach a size of $20\%$.  Above
kinematic edges, for instance above the invariant mass
$m_{\ell b}=\sqrt{m_t^2-M_W^2}$ , nonresonant contributions dominate and the
NWA fails.  The effect of off-shell effects on the extraction of the top mass
has been studied comparing the NWA to full off-shell calculations. 
Performing a template fit of the $m_{\ell b}$ distribution using partonic cross
sections,  a sizable shift of $0.8$~GeV due to off-shell
effects was reported.\cite{Heinrich:2017bqp} From the shift of the
peak position of the $m_{\ell b}$ distribution and using off-shell NLO
calculations matched to a resonance-aware parton shower,\cite{Jezo:2016ujg} a more moderate effect of
$\sim 0.1$~GeV was found.\cite{Ravasio:2018lzi}
However, larger effects were found for the {\texttt{Herwig}} shower than for
{\texttt Pythia}.  Both studies do not fully correspond to the experimental
situation so further work is needed for conclusive statements.  In
addition, other effects relevant for the interpretation of the top-mass
measurements at hadron colliders such as renormalon effects on the relation of
the pole-mass to short-distance mass definitions~\cite{Nason:2017cxd} or the
impact of parton-shower effects on the mass definition~\cite{Hoang:2018zrp}
remain controversial.
\newpage


\end{document}